# Modeling Urban Growth and Socio-Spatial Dynamics of Hangzhou, China: 1964-2010

Jian Feng, Yanguang Chen

(Department of Geography, College of Urban and Environmental Sciences, Peking University, Beijing, 100871, China; fengjian@pku.edu.cn; chenyg@pku.edu.cn)

**Abstract:** Urban population density provides a good perspective for understanding urban growth and socio-spatial dynamics. Based on sub-district data of the five times of national population censuses in 1964, 1982, 1990, 2000, and 2010, this paper is devoted to making analyses of urban growth and the spatial restructuring of population in the city of Hangzhou, China. Research methods are based on mathematical modeling and field investigation. The modeling result shows that the negative exponential function and the power-exponential function can be well fitted to Hangzhou's observational data of urban density. The negative exponential model reflect the expected state, while the power-exponential model reflects the real state of urban density distribution. The parameters of these models are linearly correlated to the spatial information entropy of population distribution. The density gradient in the negative exponential function flattened in the 1990s and 2000s is closely related to the development of suburbanization. In terms of investigation materials and the changing trend of model parameters, we can reveal the spatio-temporal features of Hangzhou's urban growth. The main conclusions can be reached as follows. The policy of reformation and opening-up and the establishment of a market economy improved the development mode of Hangzhou. As long as a city has a good social and economic environment, it will automatically tend to the optimal state through self-organization.

## 1. Introduction

Urban growth and the spatial restructuring of urban morphology based on population densities are one of the important issues of urban geography. With the introduction of post-modern mathematics such as fractal geometry and bionic mathematics such as cellular automata (Batty and Longley, 1994; Frankhauser, 1994; White and Engelen, 1993; White *et al*, 1997), the ability of quantitative depiction and method of simulated experiment of urban geography are strengthened. Fractal description and computer simulation have brought and are still bringing about a revolution in the studies on urban growth and morphology. Some traditional mathematical models on urban population densities, such as Clark's model (Clark, 1951), were re-explained in the rise of studies on spatial complexity based on the analysis of urban systems (Batty and Kim, 1992; Batty and Longley, 1994; Chen, 2008). It is noticed that, until the late 1990s, the research of modeling urban population densities in developed countries is extensive, while there is much less on cities in developing countries, especially in China (Wang and Zhou, 1999). Since the late 1990s, research on developing countries has also made some progress in this field. Three researches on urban China, with one on urban population densities in Beijing 1982-1990 by Wang and Zhou (1999), one about spatial restructuring of population towards polycentricity in Beijing 1982-2000 by Feng et al (2009), and the other one about those in Shenyang 1982-1990 by Wang and Meng (1999), brought us interesting issues and useful information about the similarities and differences between cities in China and cities in the Western Countries. Another research by Xu et al (2019) tell us about how urban population density decline over time in Chinese cities from 2001 to 2015 based on exponential model, and the authors find that initial densities and GDP growth rates influence the declining density rates. Luo et al (2019) try to model population density using a new index derived from multi-sensor image data, and then a case study is implemented using the data of Sichuan Province, China in 2015. Ren et al (2020) performed the processing of the population density grid data of Nairobi, Kenya for 2000 and 2020, to analyze population density and spatial patterns of informal settlements based on the Clark model, the Newling model and the McMillen model. In recent years, the research on urban population density and urban growth in European and American countries continues to make progress, accompanied by the improvement of methods. For example, Polinesi et al (2020) try to reflect the relationship between population trends and urbanization in Greece by simulating density effects using a local regression approach. Mariani et al (2018) illustrates a simplified procedure identifying population sub-centers over 50 years in three Southern European cities (Barcelona, Rome, Athens) by identifying metropolitan sub-centers from diachronic density-distance curves. Qiang et al (2020) provides an empirical evaluation of the classic population density functions in 382 metropolitan statistical areas (MSA) in the USA using travel times to city centers as the independent variable. Moghadam et al (2018) examine patterns of growth and change in Sydney over the time period from 1981 to 2006 taking employment density as a structural dimension, and modelling the linear and interactive effects of this dimension. The results shown that the urban population density distribution in both China and European-American countries complies with distance-decay law and can be described with the negative exponential function. The results also indicate that the main causes of urban growth in China in the 1980s, was at an incipient stage, unlike that in Western Countries which has been in full maturity. To get more credible knowledge about the comparison between population densities and related spatial dynamics of Chinese cities and those of the Western cities, it is necessary to conduct more empirical researches on population densities of urban China.

Urban growth is associated with urban form (static structure) and spatial dynamics (interaction between elements). It has always been an important topic in urban studies (Batty and Longley, 1994; Chen, 2020; Newling, 1966; Rozenfeld et al, 2008). There are two main objectives for urban growth research. One is to find the statistical laws from the macro level and describe them by mathematical modeling, and the other is to reveal the evolution mechanism from the micro level, so as to carry out empirical analysis to gain understanding and insight about city development. Anyway, scientific research include two processes: description and understanding (Gordon, 2005; Henry, 2002). The measurements for describing urban growth include population, wealth, land use area, and night light data. In early years, two central variables in the study of spatial dynamics of urban evolution are population and wealth (Dendrinos, 1992). With the development of remote sensing technology, urban land area has become an important measure (Rozenfeld *et al*, 2008; Rozenfeld *et al*, 2011; Schneider et al, 2004). Recent years, night lights provide new observational data for cities (Long and Chen, 2019; Zeng *et al*, 2011). Research perspectives involve time (processes), space (patterns), and hierarchy (size distribution), and are associated with three basic laws of geography: allometric growth law, distance-decay law, and rank-size law. Conventional mathematical modeling and quantitative analyses of urban growth are based on characteristic length. Due to development of fractal geometry, fractal dimension and scaling exponent become important parameters for characterizing urban growth (Batty and Longley, 1994; Feng and Chen, 2010; Frankhauser, 1994; Lu and Tang, 2004; Makse et al, 1995; Makse *et al*, 1998; Manrubia and Zanette, 1999; Peterson, 1996). Using time series of fractal dimension, we can build fractal dimension increase models to exploring spatial dynamics of urban evolution (Man and Chen, 2020).

Although urban growth research can be made by using various new technologies and methods, two facts are undeniable. First, population as a central variable for urban analysis is irreplaceable. City is the space and place for human life. Population represents the first dynamics of urban growth (Arbesman, 2012). Second, investigation as a basic method for urban studies is irreplaceable. Seeing is believing. Without on-the-spot investigation, it is difficult to understand the truth of a city's development (Zhou, 1995). Based on field research and systematic census data, this paper is devoted to research the distribution of population density in Hangzhou with the help of mathematical modeling and empirical analysis. Through spatio-temporal evolution of urban density distribution, we can explore the socio-spatial dynamics of Hangzhou's urban growth. The purpose is to reveal the law of urban growth in China from the perspective of population evolution. It also provides a reference for the comparison of urban development between China and the Western countries. The rest parts are organized as follows. In Section 2, the basic and new models for describing urban density are reviewed and explained, and data and method are clarified. In Section 3, empirical analyses are made for spatio-temporal evolution of Hangzhou's population density distribution patterns, and the spatial restructuring process of population are analyzed. In Section 4, several related questions are discussed. Finally, the discussion is concluded by summarizing the main points of view.

## 2. Models and methods

### 2.1. Mathematical models

The mathematical models of urban population density can be traced back to the early 1950s. A number of functions are proposed to characterize urban density distribution (Batty and Longley, 1994; Chen, 2010; Mcdonald, 1989). By making statistical analysis of more than twenty cities, Clark (1951) found that urban population density tends to decline in a negative exponential fashion with increasing distance from the city center. Clark's model can be expressed as

$$\rho(r) = \rho_0 e^{-br}, \tag{1}$$

where $r$ is distance, $\rho(r)$ is the population density at distance r from the city center, $\rho_0$ is the density at the city center ($r=0$), $b$ is the rate at which population density declines with distance r. Clark's law represents the classic model for urban population density distribution. Several years after that, geography entered the stage of Quantitative Revolution. During the period, geographers studied theoretical models with immense zeal. In the early 1960s, two new models of population density were put forward: one is a normal model by Sherratt (1960) and Tanner (1961), and the other is an inverse power model by Smeed (1963). Sherratt-Tanner model can be expressed as

$$\rho(r) = \rho_0 e^{-br^2}, \tag{2}$$

which is a form of Gaussian function. The model indicates that in the area far away from the city center, population density based on this model decreases much faster than that in Clark's model. Smeed's model can be express in the following form

$$\rho(r) = Kr^{-\alpha}, \tag{3}$$

where $K$ is the constant of proportionality, and $\alpha$ is the parameter on distance. It is easy to find that Smeed's model is not defined where $r = 0$. However, Smeed's model cannot fit most of the distribution of population density. Parr (1985) once pointed out that the negative exponential function is more appropriate for describing density in the urban area, while the inverse power function is more appropriate to the urban fringe and hinterland. Later, Longley and Mesev (1997) argued that the spatial distribution of population density should be depicted from two kinds of scales: Smeed's model can fit the spatial distribution of population density well when r is smaller, while urban population density decreases faster than the model's expectation when r is larger.

Sometimes, all the above models cannot reflect urban density distribution well, and more complicated models seem to be necessary. Replacing the linear equation in Clark's model with a quadratic function, Newling (1969) suggested a quadratic exponential model as follows

$$\rho(r) = \rho_0 e^{br-cr^2}, \tag{4}$$

where $b$ and $c$ are parameters. Because the exponential term in Newling's model contains a multinomial equation, Newling's model can fit realistic data of population density preferably. Newling's model was treated as a more satisfactory one than that of Clark in describing the spatial distribution of population density. Based on Newling's model, a series of theories such as the crater effect, the tide wave of metropolitan expansion, and the stages of urban evolution are all developed at that time (Latham and Yeates, 1970; Cadwallader, 1996; King and Golledge, 1978). Chen (2010) proposed a power-exponential model of urban density, which incorporates Clark's model and Sherratt-Tanner's model directly. The model can be expressed as

$$\rho(r) = \rho_0 e^{-br^\sigma}, \tag{5}$$

where $\sigma$ is a parameter which reflects the changes of information entropy of urban population distribution. It can be treated as a latent scaling exponent. We will explain this in the following text. It can be seen that when σ=1, equation (5) will reduce to Clark's model, and that when σ=2, equation (5) will change to Sherratt-Tanner's model. Moreover, probably the lognormal distribution is also a choice (Parr and O'Neill, 1989). The model is expressed as

$$\rho(r) = \rho_0 e^{-b(\ln r)^2}. \tag{6}$$

All the above-mentioned models will be tested by using observed data in this research.

For a long time, the models of urban population density were empirical models rather than theoretical models. If and only if a model is derivable from general principles, it will lead us to its underlying rationale. Around the 1970s, Clark's model were theoretically derived from a utility-maximizing model by assuming a unitary value of the price elasticity of housing services (Muth, 1969; Mills, 1972). Another attempt is to derive the Clark model by means of entropy-maximizing method. Based on spatial interaction models for traffic distribution, a negative exponential decay model was derived by Bussiere and Snickars (1970). The process was regarded as indirect derivation of Clark's model through entropy-maximization idea (Batty and Longley, 1994). By postulating urban cell system, Chen (2008) derived Clark's model gracefully by means of entropy-maximizing principle. The direct derivation of Chen (2008) is simpler and clearer than that of Bussiere and Snickars (1970). Entropy-maximization suggests the best balance between the equity of individuals and the efficiency of the whole (Chen, 2012). Entropy-maximization may be associated with urban sustainable development of cities (Chen, 2010).

## 2.2. Study area and data

In this part, we first present a brief overview of the study area, Hangzhou Municipality. We then explain the method to deal with the census data, and the data with which we conduct our research. Hangzhou, as the capital of Zhejiang Province in the east of China, is famous for its scenic beauty. In 2000, the city region of Hangzhou included a mostly urbanized municipality and rural region. The urban municipality comprises six districts, i.e., *qu*, including Shangcheng, Xiacheng, Jianggan, Gongshu, Xihu, and Binjiang, and the rural region consists of seven counties, that is, *xian*, including Xiaoshan, Yuhang, Fuyang, Linan, Tonglu, Jiande and Chunan, under its administration. From 2000 to 2019, Xiaoshan, Yuhang, Fuyang and Linan also became districts from counties under the control of the government. Considering that the main body of urbanized region in Hangzhou is still made up of the six districts in 2000, which is called Hangzhou Municipality, including Shangcheng, Xiacheng, Jianggan, Gongshu, Xihu, and Binjiang, we take Hangzhou Municipality as the study area. Hangzhou Municipality includes the urban core (Figure 1), including Shangcheng and Xiacheng, and the inner suburb, including Jianggan, Gongshu, Xihu, and Binjiang. The outer suburb includes Xiaoshan, Yuhang, Fuyang, Linan, Tonglu, Jiande and Chunan (Feng and Zhou, 2005).

In 1964 (the second census), 1982 (the third census), 1990 (the forth census), 2000 (the fifth census) and 2010 (the sixth census), the municipality contained a population of 1,071.0, 1,348.1, 1,647.5, 2,451.1 and 3,560.4 thousand respectively. According to the area of Hangzhou Municipality in 2000, the population density in each year was 1516.4, 1908.8, 2332.6, and 3470.5 and 5041.1 persons per square meter (later abbreviated as p/km2) respectively. From 1964 to 2010, the administrative divisions of Hangzhou sub-district level (*jiedao*, *zhen*, *xiang*) have changed, which makes it difficult to compare the data before and after a certain period. For the convenience of the research, some sub-districts are merged, which is called sub-district unit. The municipality of Hangzhou was composed of 44 sub-district units in 1964 and 1982, 47 ones in 1990, 50 ones in 2000 and 44 ones in 2010. A sub-district unit is an irregular small zone, which form a zonal system of Hangzhou (Figure 1). Hangzhou had been the capital of China in the ancient times, especially in the South Song Dynasty. In modern times, Hangzhou is one of the important cities in the Yangtze Delta and has a close relationship with Shanghai. Therefore Hangzhou will provide a good case for empirical studies of population densities in urban China.

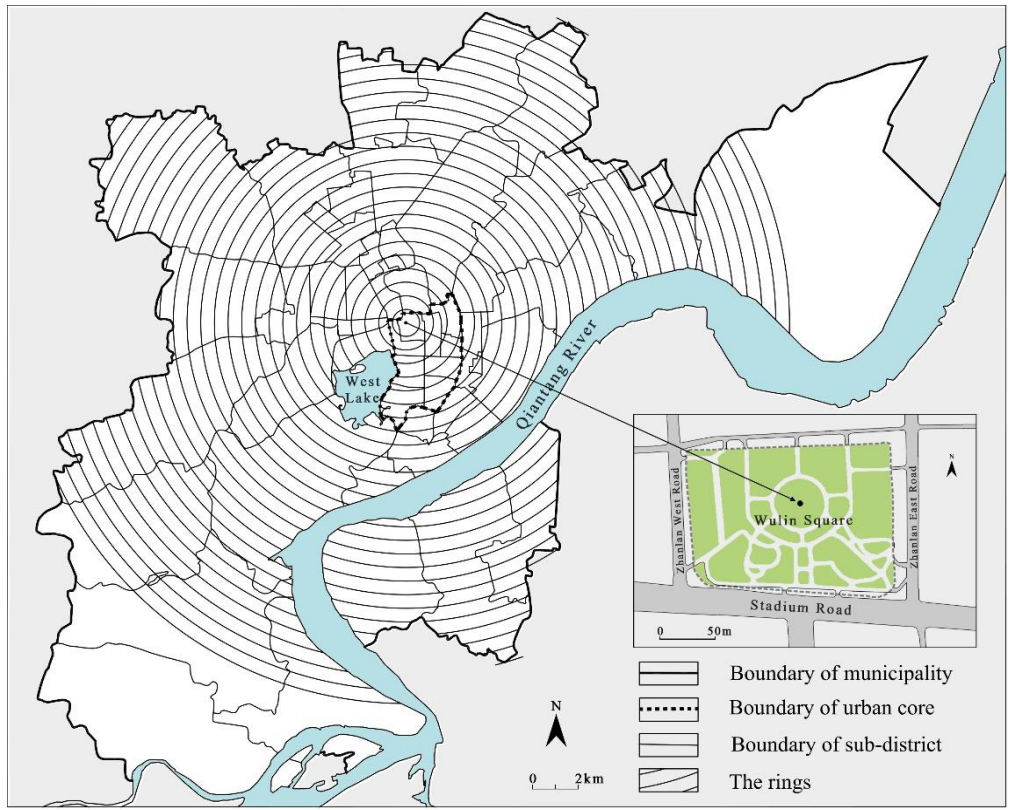

**Figure 1.** A map of zonal system indicating sub-districts of Hangzhou and corresponding system of concentric circles. Note: For urbanized area, a sub-district is termed *jiedao* in Chinese, and for rural area, sub-district is termed *Xiang* and *Zhen*.

## 2.3. Data processing methodology

The firsthand data come from the second, third, fourth, fifth and sixth censuses in 1964, 1982, 1990, 2000, and 2010, respectively. In order to study urban growth and the spatial-temporal evolution of urban geographic system, the data of population by sub-districts are transformed into those by spatial distribution. The data are dealt with according to the following four steps. (1) Adjust the boundaries of each sub-districts according to historical materials and records. We obtain the map with spatial boundaries of sub-districts of Hangzhou in the four years from various sources of local governments and numerous field trips. To get the population of sub-districts whose boundaries are definite, some sub-districts are merged, according to the changes of the boundaries of sub-districts in 1964-2010. (2) Calculate population density of each sub-district in the five years (i.e., 1964, 1982, 1990, 2000, and 2010). The boundaries of sub-districts form a zonal system of each year (Batty and Longley, 1994; Chen, 2020). From the map of zonal systems, we can obtain the data of area of each sub-district in the five years. At last, the data of population density of each sub-district in the five years are calculated according to sub-district population and area. (3) Determine the city center of Hangzhou. The central location is identified as Wulin Square. The main bases are as follows: (a) The busiest area (esp. that of business and trade) of the city is located around Wulin Square, and as a result some scholars think it is the "old CBD" of Hangzhou (Kong and Dong, 2006; Zhang, 2012; Chu, 2018), because since 2001, Hangzhou has built a modern CBD in Qianjiang New Town on the North Bank of Qiantang River. There is no doubt that the newly built CBD (Qianjiang new town) cannot become the urban center of Hangzhou, no matter from the geographical location or from the actual function; (b) The population densities of several sub-districts near Wulin Square are of the highest level constantly within several decades; (c) Wulin Square has been regarded as the center of modern and contemporary activities of Hangzhou city (Kong and Dong, 2006); and (d) Wulin Square is the centroid of Hangzhou municipality. So the center of Wulin Square is regarded as the approximating position of city center of Hangzhou. (4) Compute the average population density by rings. That is to say, the population density based on zonal systems is turned into that based on concentric circles.

The spatial series of urban population density in Hangzhou are the results of spatial weighted average. The methods of transforming zone-based population density data in irregular sub-districts into ring-based population density data in regular circular banded zones should be explained. The process of data extraction are as follows.

**Step 1: calculate the population density of each sub-district.** Suppose there are $n$ sub-districts in Hangzhou in a given year. The sub-districts can be numbered as $j$=1, 2, 3,…, $n$. If the population size and area of the $j$th sub-district are $P_j$ and $A_j$, respectively, then the population density is

$$\rho_j = \frac{P_j}{A_j}. \tag{7}$$

This step is very simple can be realized by MS Excel.

**Step 2: define circular ring zones.** Taking the center of Wulin Square as the center of circles and $l_i$ as radiuses, we can draw a series of concentric circles. The difference of radius is $\Delta l = l_{i+1} - l_i = 0.6$. Two adjacent circles form a circular ring zone. The radius of the $i$th circular ring zone can be expressed as

$$l_i = 0.6(i - 1/2), \tag{8}$$

where $i$=0, 2, …, 26 (unit: km). The first circle is actually a point, and $i$=0 represents city center point. Thus the main part of Hangzhou municipality are divided into 26 circular ring zones (see Figure 1).

**Step 3: compute the intersectional areas of circular ring zones and administrative sub-districts.** As indicated above, a sub-district is an irregular small region, while a circular ring zone is a regular banded region. If two adjacent circles cut a sub-district, there will be an intersectional area between the corresponding regular ring zone and the irregular sub-district. For the $i$th circular zone and the $j$th sub-district, the intersectional area can be expressed as

$$S_{ij} = \rho_j \bigcap (0.6^2 \pi ((i+1)^2 - i^2)), \tag{9}$$

where $S_{ij}$ denotes the intersectional area, and $\bigcap$ means to take the intersection of two sets ($i=1,2,…,26$; $j=1,2, …,n$; $0 \leq S_{ij} \leq S_j$). This step is the most complicated one, which can be realized with the help of Arc GIS technology.

**Step 4: defining population density weights.** The intersectional areas of the $i$th circular ring zone and the $n$ sub-districts can serve as spatial weights for circular density computation. The formula is

$$W_{ij} = S_{ij} / \sum_{j=1}^{n} S_{ij}, \tag{10}$$

where $W_{ij}$ denotes spatial weight for average population density.

**Step 5: calculate the average density of circular ring zones.** The average population density of the $i$th circular ring zone can be calculated by the following formula

$$\rho_i = \sum_{j=1}^{n} W_{ij} \rho_j, \tag{11}$$

where $\rho_i$ is the average population density of the $i$th circular ring zone. Using the above-illustrated method, step by step, we worked out the average population density datasets of Hangzhou in 1964, 1982, 1990 and 2000 (Table 1). As a supplement, the data of sub-district population density is also used to analyze the spatial restructuring of population.

**Table 1.** Datasets of urban population density and distance by ring zones in Hangzhou in 1964, 1982, 1990, 2000, and 2010

| The order of rings ($i$) | Distance (km) | Population density (p/km²) | | | | |
|---|---|---|---|---|---|---|
| | | 1964 | 1982 | 1990 | 2000 | 2010 |
| 1 | 0.3 | 24,131 | 29,540 | 29,928 | 28,184 | 26,635 |
| 2 | 0.9 | 18,966 | 22,225 | 26,634 | 26,821 | 25,419 |
| 3 | 1.5 | 16,282 | 18,957 | 22,262 | 24,621 | 22,702 |
| 4 | 2.1 | 16,007 | 19,232 | 21,612 | 23,176 | 20,923 |
| 5 | 2.7 | 13,052 | 15,439 | 17,290 | 18,910 | 19,466 |
| 6 | 3.3 | 8,259 | 9,929 | 12,478 | 16,911 | 18,830 |
| 7 | 3.9 | 5,798 | 7,026 | 9,896 | 14,522 | 16,594 |
| 8 | 4.5 | 2,626 | 3,461 | 5,560 | 10,829 | 12,428 |
| 9 | 5.1 | 2,143 | 2,807 | 4,180 | 7,282 | 9,226 |
| 10 | 5.7 | 2,142 | 2,689 | 3,923 | 6,200 | 7,996 |
| 11 | 6.3 | 2,185 | 2,566 | 3,516 | 5,644 | 7,363 |
| 12 | 6.9 | 1,438 | 1,693 | 2,197 | 4,297 | 6,487 |
| 13 | 7.5 | 1,083 | 1,371 | 1,796 | 3,806 | 5,863 |
| 14 | 8.1 | 967 | 1,256 | 1,634 | 3,153 | 5,260 |
| 15 | 8.7 | 842 | 1,114 | 1,442 | 2,683 | 4,830 |
| 16 | 9.3 | 848 | 973 | 1,265 | 2,354 | 4,509 |
| 17 | 9.9 | 818 | 1,051 | 1,163 | 2,028 | 4,200 |
| 18 | 10.5 | 812 | 1,051 | 1,143 | 1,828 | 4,062 |
| 19 | 11.1 | 807 | 1,051 | 1,160 | 1,651 | 3,846 |
| 20 | 11.7 | 625 | 979 | 1,093 | 1,581 | 3,788 |

| 21 | 12.3 | 691 | 901 | 1,006 | 1,490 | 3,673 |
| 22 | 12.9 | 575 | 870 | 972 | 1,465 | 3,543 |
| 23 | 13.5 | 532 | 666 | 817 | 1,278 | 2,953 |
| 24 | 14.1 | 381 | 487 | 679 | 1,033 | 2,630 |
| 25 | 14.7 | 369 | 489 | 582 | 958 | 2,351 |
| 26 | 15.3 | 375 | 456 | 563 | 882 | 2,125 |

## 3. Empirical results

### 3.1. Modeling population densities in Hangzhou

Mathematical modeling is a good approach to deep researching urban form and growth. The mathematical expression of a model reflects the macro structure of a city, while the parameter values of the model reflect the dynamic characteristics of the micro elements of the city. In scientific research, mathematical models can be classified into two categories: mechanism models and parametric models (Su, 1988). Accordingly, there exist two methods for establishing mathematical models: analytical method and experimental method (Zhao and Zhan, 1991). This paper is devoted to making parametric models of urban growth and form based on population density distribution by using experimental method. The modeling process of experiment method for parameter models is to select one or two most possible functions from a set of possible functions by comparison and analysis. However, the ideas from the analytical method for mechanistic models can be used to make complement judgement.

The functions discussed in the preceding section can be fitted to the data of average population densities in 1964, 1982, 1990, 2000 and 2010. The method of parameter estimation is the ordinary least square (OLS) regression. With the help of $r$-$D(r)$ scatterplots, we can select the possible models. We test four simple functions, including the linear one, the exponential one, the logarithmic one and the power one, and three more complicated functions, including the lognormal one, the power-exponential one and the quadratic exponential one. The expression of each function and the result of regressions are listed in Tables 2 and 3. Apparently, using the data of average urban population density brings good results of regression. For example, with the exception of linear model, all the coefficient determinations $R^2$ of the other models are above 0.8, with most of them above 0.9.

Table 2. The parameter estimation results of the models population densities in Hangzhou (1964-2010)

| Year | Type of model | Expression of model | $a$ | $b$ | $R^2$ | $F$ |
|---|---|---|---|---|---|---|
| 1964 | Linear | $\rho(r) = a-br$ | 14,019.571 | 1,192.114 | 0.631 | 41.000 |
|  | Exponential | $\rho(r) = a\exp(-br)$ | 16,429.413 | 0.281 | 0.907 | 234.169 |
|  | Logarithmic | $\rho(r) = a-b\ln r$ | 16,799.930 | 6,860.880 | 0.909 | 240.407 |
|  | Power | $\rho(r) = ar^{-b}$ | 19,116.626 | 1.329 | 0.886 | 186.192 |
|  | Lognormal | $\rho(r) = a\exp(-b(\ln r)^2)$ | 16,430.129 | 0.550 | 0.920 | 276.889 |
|  | Power-exponential | $\rho(r) = a\exp(-br^\sigma))$ | 85,622.987 | 1.535 | 0.958 | 546.829 |
| 1982 | Linear | $\rho(r) = a-br$ | 16,781.830 | 1,420.395 | 0.630 | 40.901 |

|  | Exponential | $\rho(r) = a\exp(-br)$ | 19,493.085 | 0.272 | 0.904 | 225.927 |
|---|---|---|---|---|---|---|
|  | Logarithmic | $\rho(r) = a - b\ln r$ | 20,143.212 | 8,202.296 | 0.915 | 256.917 |
|  | Power | $\rho(r) = ar^{-b}$ | 22,712.030 | 1.294 | 0.887 | 188.962 |
|  | Lognormal | $\rho(r) = a\exp(-b(\ln r)^2)$ | 19,455.030 | 0.533 | 0.915 | 259.698 |
|  | Power-exponential | $\rho(r) = a\exp(-br^\sigma)$ | 113,984.359 | 1.637 | 0.957 | 527.971 |
| **1990** | Linear | $\rho(r) = a - br$ | 19,408.637 | 1,626.398 | 0.681 | 51.262 |
|  | Exponential | $\rho(r) = a\exp(-br)$ | 24,583.475 | 0.275 | 0.931 | 323.755 |
|  | Logarithmic | $\rho(r) = a - b\ln r$ | 22,658.850 | 9,051.841 | 0.918 | 269.276 |
|  | Power | $\rho(r) = ar^{-b}$ | 27,182.362 | 1.274 | 0.872 | 162.981 |
|  | Lognormal | $\rho(r) = a\exp(-b(\ln r)^2)$ | 24,618.028 | 0.539 | 0.946 | 418.267 |
|  | Power-exponential | $\rho(r) = a\exp(-br^\sigma)$ | 80,862.588 | 1.140 | 0.966 | 687.942 |
| **2000** | Linear | $\rho(r) = a - br$ | 21,927.176 | 1,757.632 | 0.784 | 87.323 |
|  | Exponential | $\rho(r) = a\exp(-br)$ | 30,786.603 | 0.250 | 0.972 | 826.901 |
|  | Logarithmic | $\rho(r) = a - b\ln r$ | 24,239.823 | 9,100.723 | 0.915 | 259.055 |
|  | Power | $\rho(r) = ar^{-b}$ | 30,760.150 | 1.108 | 0.829 | 116.691 |
|  | Lognormal | $\rho(r) = a\exp(-b(\ln r)^2)$ | 30,588.663 | 0.489 | 0.979 | 1,140.566 |
|  | Power-exponential | $\rho(r) = a\exp(-br^\sigma)$ | 47,053.952 | 0.523 | 0.981 | 1,226.034 |
| **2010** | Linear | $\rho(r) = a - br$ | 21,785.021 | 1,571.545 | 0.824 | 112.309 |
|  | Exponential | $\rho(r) = a\exp(-br)$ | 26,447.238 | 0.172 | 0.960 | 583.084 |
|  | Logarithmic | $\rho(r) = a - b\ln r$ | 23,501.138 | 7,937.438 | 0.915 | 257.357 |
|  | Power | $\rho(r) = ar^{-b}$ | 26,900.863 | 0.770 | 0.841 | 127.359 |
|  | Lognormal | $\rho(r) = a\exp(-b(\ln r)^2)$ | 26,296.234 | 0.335 | 0.966 | 683.762 |
|  | Power-exponential | $\rho(r) = a\exp(-br^\sigma)$ | 39,012.813 | 0.430 | 0.974 | 906.916 |

**Notes**: In this paper, all parameter estimates are based on 14 decimal places, but only 3 decimal places are shown. In the power-exponential model, the restraint parameter, that is, latent scaling exponent, $\sigma$, is equals to 0.475 in 1964, 0.450 in 1982, 0.550 in 1990, 0.756 in 2000, and 0.699 in 2010, respectively.

**Table 3.** The regression results for a quadratic exponential model of population density in Hangzhou (1964-2010)

| Year | Quadratic exponential model | Transformed by logarithm | $R^2$ |
|---|---|---|---|
| **1964** | $\rho(r) = 35,383.469 e^{-0.576r + 0.019r^2}$ | $\ln\rho(r) = 10.474 - 0.576r + 0.019r^2$ | 0.974 |
| **1982** | $\rho(r) = 41,150.856 e^{-0.560r + 0.018r^2}$ | $\ln\rho(r) = 10.625 - 0.560r + 0.018r^2$ | 0.971 |
| **1990** | $\rho(r) = 47,120.711 e^{-0.525r + 0.016r^2}$ | $\ln\rho(r) = 10.760 - 0.525r + 0.016r^2$ | 0.982 |
| **2000** | $\rho(r) = 42,744.508 e^{-0.376r + 0.008r^2}$ | $\ln\rho(r) = 10.663 - 0.376r + 0.008r^2$ | 0.988 |
| **2010** | $\rho(r) = 34,031.936 e^{-0.269r + 0.006r^2}$ | $\ln\rho(r) = 10.435 - 0.269r + 0.006r^2$ | 0.981 |

The parameter estimation results suggest the possible models for Hangzhou's population density distribution. As far as the four simple functions are concerned, it is easy to exclude the linear model with the worst goodness of fit, according to the results of regressions (Table 2). Generally, speaking, urban population density distribution follow the distance decay law. Linear function is not one of the alternative models of distance decay patterns in geography (Haggett *et al*, 1977; Taylor, 1983; Zhou, 1995). The goodness of fit of the exponential one is a bit worse than that of the logarithmic one in 1964 and 1982, while the former is better than the latter since 1990, especially in 2000 and 2010. So totally speaking, the Clark's model occupies a dominant position in modeling population densities of Hangzhou in the 1990s. In particular, the exponential model bears good theoretical basis in describing urban population density distribution (Bussiere and Snickars, 1970; Chen, 2008). It is not difficult to find that the power model (Smeed's model) does not dominate in the analysis of Hangzhou's population density. As a result, it is apparent that the density distribution of urban population is not fractal. The results of lognormal model and that of power-exponential model are often close to one another, but the lognormal model has not enough theoretical basis. Although the lognormal model can be fitted better to the observational data in some year, it cannot be the total tendency. The coefficient of determination ($R^2$) is an important basis for model selection, but not the only one (Shelberg *et al*, 1982). When we extend the curve of the quadratic exponential model fitted by Hangzhou's data to some position in suburban, population density increases again and reaches a certain value which is much higher than that in the urban center. This phenomenon can be treated as over fitting (Silver, 2012). The result is absurd for it does not accord with the reality, so this model is excluded. In fact, there is an essential distinction in model structure between Newling's model and the quadratic exponential models regressed by the data of Hangzhou. In Newling's model, the coefficient of first order term is positive, and that of quadratic order term is negative, while in the quadratic exponential models regressed by Hangzhou's data in the five years, the coefficients of first order term are negative, and those of quadratic order term are positive. The difference in model structure shows that we can't get a model like that of Newling, and that there isn't any crater and tidal wave of expansion in the evolution of spatial distribution of population density in Hangzhou within the past several decades. The difference in model structure also decides the above-mentioned absurd result.

It is difficult to choose a mathematical model for complex systems. Only through comprehensive analysis can we choose the appropriate mathematical expressions. Albeit the goodness of fit of the quadratic exponential model, we also exclude it (see Table 3). Compared with the four simple models, the lognormal model and the power-exponential model are more careful ones. As a result, the goodness of fit of the two complicated functions is better than the four simple ones. Except for 1990, the power-exponential model has better goodness if fit to Hangzhou's population density than lognormal model. Based on the above analyses, two models can be employed to analyze Hangzhou's population density distribution. One is Clark's negative exponential model, and the other is power-exponential model. The latter is the generalized result of the former. Clark's model can be derived by using the principle of entropy maximization. Its theoretical rationale is clear (Batty and Longley, 1994; Chen, 2008). According to the ideas from analytical method for mechanistic models, we select Clark's model to describe the expected form of urban population density. On the other hand, the power-exponential model shows good effect of fitting to population distribution of Hangzhou. In terms of the ideas from the experimental method for parameter models, we choose this model to depict the real form of Hangzhou population density. In fact, geography is involved to two worlds: real world and ideal world (Tang, 2009). Clark's model is mainly suitable for the urban density in the ideal world, while the power-exponential model is more suitable for the urban density in the real world. The values of the hidden scaling exponent of the power-exponential model, $\sigma$, reflect the deviation degree of the real state of urban population density distribution to the ideal state that is based on entropy maximization.

### 3.2. Urban growth reflected by model parameters

The changes of some parameters in the urban density models may indicate the features of urban growth in Hangzhou for about half a century since 1964. First of all, the parameter σ of the power-exponential model reflects the tendency of changes of information entropy of urban geographic system essentially. Entropy is important measure for spatial analysis of Hangzhou's urban growth (Chen *et al*, 2017). In fact, entropy theory is very important in theoretical and empirical studies of geographical systems (Batty, 2010; Batty *et al*, 2014; Wilson, 2010). In order to make it clear, we estimate the information entropy of the spatial distribution of population in Hangzhou city. The core, inner suburb and outer suburb constitute the three big zones of Hangzhou city. Population of each city zone is displayed in Table 4. The proportion of each zone's population to city total can be regarded as the probability of spatial distribution of population. So we can use Shannon's formula of Shannon entropy to estimate the information entropy of population distribution in Hangzhou. The formula is as below:

$$H = -\sum_{i=1}^{N} p_i \log_2 p_i, \quad (12)$$

where $H$ refers to Shannon's information entropy, $p_i$ denotes the probability of zone $i$ ($i$=1, 2, 3), $N$ =3 is the number spatial units. It is easy to find that the information entropy of Hangzhou's population distribution of the three big zone keeps increasing but slightly fluctuates, indicating that the urban geographic system of Hangzhou is becoming more complex and orderly than ever. Moreover, the changes of information entropy are consistent with those of parameter σ of the power-exponential model, because there is a linear relationship between them, that is, $\sigma$=3.106$H$-2.833 ($R^2$=0.980). So the changes of parameter σ reflect the tendency of changes of information entropy of urban geographic system.

**Table 4.** Changes of some parameters of the density models and information entropy of Hangzhou's urban system

| Item | 1964 | 1982 | 1990 | 2000 | 2010 |
|---|---|---|---|---|---|
| **Gradient of the negative exponential model** | 0.281 | 0.272 | 0.275 | 0.250 | 0.172 |
| **Characteristic radius $r_0$ (km)** | 3.564 | 3.671 | 3.641 | 3.996 | 5.824 |

| Latent scaling exponent σ of the power-exponential model | 0.475 | 0.450 | 0.550 | 0.756 | 0.699 |
|---|---|---|---|---|---|
| Characteristic radius $r_0^*$ (km) | 1.946 | 1.971 | 2.339 | 3.414 | 5.580 |
| Gap between two characteristic radii $\Delta r$ (km) | 1.619 | 1.700 | 1.302 | 0.582 | 0.244 |
| Population of Hangzhou's core (person) | 430,302 | 462,390 | 407,536 | 341,633 | 298,162 |
| Population of Hangzhou's inner suburb (person) | 640,705 | 885,754 | 1,239,930 | 2,109,453 | 3,262,229 |
| Population of Hangzhou's outer suburb (person) | 3,146,324 | 3,912,328 | 4,184,668 | 4,426,893 | 5,139,982 |
| Population of Hangzhou's total city (person) | 4,217,331 | 5,260,472 | 5,832,134 | 6,877,979 | 8,700,373 |
| Information entropy of population distribution of Hangzhou (bit) | 1.064 | 1.059 | 1.087 | 1.147 | 1.146 |

**Note**: The first characteristic radius, $r_0$, is based on Clark's model, which reflects the ideal state of population density decay. The second characteristic radius, $r_0^*$, is based on the power-exponential model, which reflects the real state of population density distribution.

The entropy-maximizing defined at the macro level is consistent with the utility-maximizing defined at the micro level. In fact, scholars finished the demonstration that disaggregate models of individual resource allocation in space based on utility-maximizing were consistent with models of spatial interaction based on ideas from entropy-maximizing (Batty, 2000). Entropy-maximizing is a goal or a kind of tendency in the evolution of self-organizing system, while the information entropy of urban geographic system in reality may not be but can tend to be the maximum. From this point of view, the tendency of the change of parameter σ in the power-exponential model reflects the optimization in the process of self-organizing evolution of each function unit of Hangzhou. For Hangzhou city, the latent scaling exponent σ in the power-exponential model fluctuates, increases and approaches 1, namely, from 0.475 in 1964, to 0.450 in 1982, to 0.550 in 1990, 0.756 in 2000, and then to 0.699 in 2010. Correspondingly, the spatial information entropy values are 1.064, 1.059, 1.087, 1.147, and 1.146 bit in these years. The latent scaling exponent is highly positive linear correlation to the spatial information entropy. This suggests that the power-exponential distribution of urban population density evolves into the ideal Clark model with the lapse of time. The parameter values and its change suggests constrained entropy-maximizing process of Hangzhou's urban evolution.

Next, we may analyze the features of urban growth in Hangzhou by some parameters in Clark's model. Clark's law is the ideal form of the power-exponential model and some analytical functions of the latter can be realized directly by the former. Moreover, Clark's model also describe Hangzhou's population density well. In the negative exponential model, the parameter *a* refers to the theoretic population density of city center, and the absolute value of the parameter *b* represents the gradient of population density decay. Figure 2 provides us the comparison of the gradients in Hangzhou in 1964, 1982, 1990, 2000, and 2010.

Totally speaking, the fact tells us that the gradient tends to flatten but fluctuates with the lapse of time, for the gradient declined from 0.281 in 1964 to 0.272 in 1982, and slightly rose to 0.275 in 1990, and then declined to 0.250 in 2000, and 0.172 in 2010. The gradient coefficient is linear negative correlation to the spatial information entropy. The change of the gradients before the 1990s is inappreciable, while that after the 1990s is distinct, especially in 2000s. In Figure 2, the three lines in 1964, 1982 and 1990 are almost parallel, while the lines in 2000 and 2010 slopes more gently than the former three ones, indicating the gradient of population density become flatter than ever since the 1990s. As far as population densities at the city center are concerned, its theoretical values ($a$) increased from 16,429 p/ km² in 1964, to 19,493 p/ km² in 1982, to 24,583 p/ km2 in 1990, to 30,787 p/ km² in 2000, and then 26,447 p/ km² in 2010. Its actual values increased first, and then began to decline during the period from 1990 to 2000 (24,131 p/ km2 in 1964, 29,540 p/ km² in 1982, 29,928 p/ km² in 1990, 28,184 p/ km² in 2000, and 26,635 p/ km² in 2010). Therefore, in other words, the actual value of population density in the center of the city has been declining since 2000, but the theoretical value has been declining since 2010. Obviously, the decline time of the theoretical value is later than that of the actual value. In Figure 2b, the line in 2010 not only becomes more flat than that in 2000, but also the highest value of the line in 2010 is lower than that in 2010, which is a reflection of the model structure. In sum, the changes of parameters in the power-exponential model and the negative exponential model reflects some characteristics of urban growth in Hangzhou in the period from 1964 to 2010. They are closely related to the development of urban economy and the concentration and decentralization of population. The explanation will be given in the following section.

The spatial feature of urban population density distribution can be reflected by characteristic radius. Both Clark's model and the power-exponential model can give the estimated value of urban characteristic radius of population distribution. Where Clark's model is concerned, the reciprocal of the density gradient is just the characteristic radius of urban population distribution (Chen, 2008), and the formula is

$$r_0 = \frac{1}{b}. \qquad (13)$$

Using this formula, we can easily calculate the characteristic radius of Hangzhou's population distribution: 3.564 km in 1964, 3,671 km in 1982, 3.641 km in 1990, 3.996 km in 2000, and 5.824 in 2010. Clark's model reflects the ideal state of urban population density distribution, thus $r_0$ can be treated as ideal or expected characteristic radius. Based on the power-exponential model, the characteristic radius can be estimated by the following relation (Chen, 2010)

$$r_0^* = (\frac{1}{\sigma b})^{1/\sigma}. \qquad (14)$$

By means of this formula, we can calculate another type of characteristic radius of Hangzhou's population distribution: 1.946 km in 1964, 1,971 km in 1982, 2.339 km in 1990, 3.414 km in 2000, and 5.580 in 2010. The power-exponential model reflects the real state of urban population density distribution, thus $r_0^*$ can be regarded as actual characteristic radius. Urban characteristic radius reflects the average distance of activities of urban residents, and it depends on the development of technology and means of transportation. The different between the two characteristic radii can be computed by

$$\Delta r = r_0 - r_0^* = \frac{1}{b} - (\frac{1}{\sigma b})^{1/\sigma}, \tag{15}$$

where reflects the gap between the real state and ideal state of urban population density distribution. The gaps in different years are as follows: 1.619 km in 1964, 1,700 km in 1982, 1.302 km in 1990, 0.787 km in 2000, and 0.752 in 2010. It can be seen that with the development of the city of Hangzhou, the actual characteristic radius is more and more close to the expected characteristic radius. Hangzhou city development has the trend of spatial optimization by self-organization.

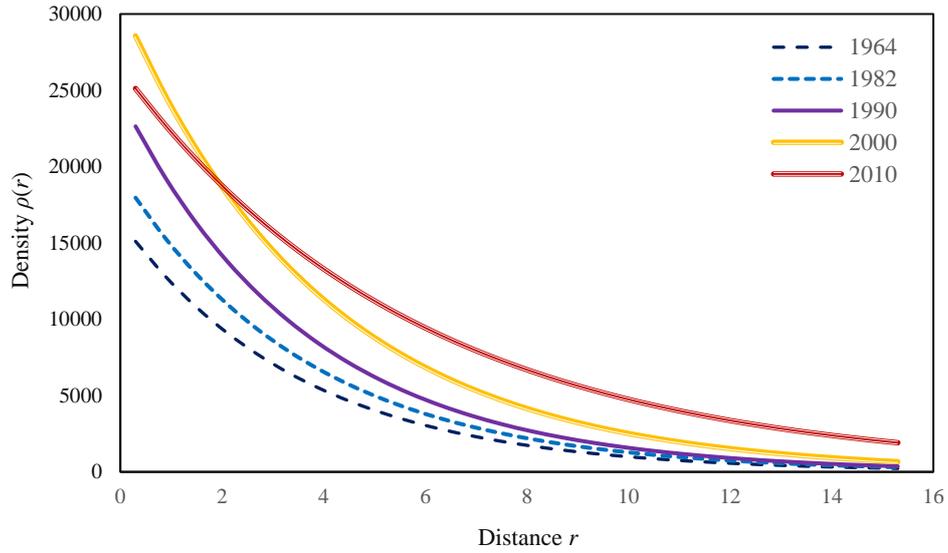

(a)

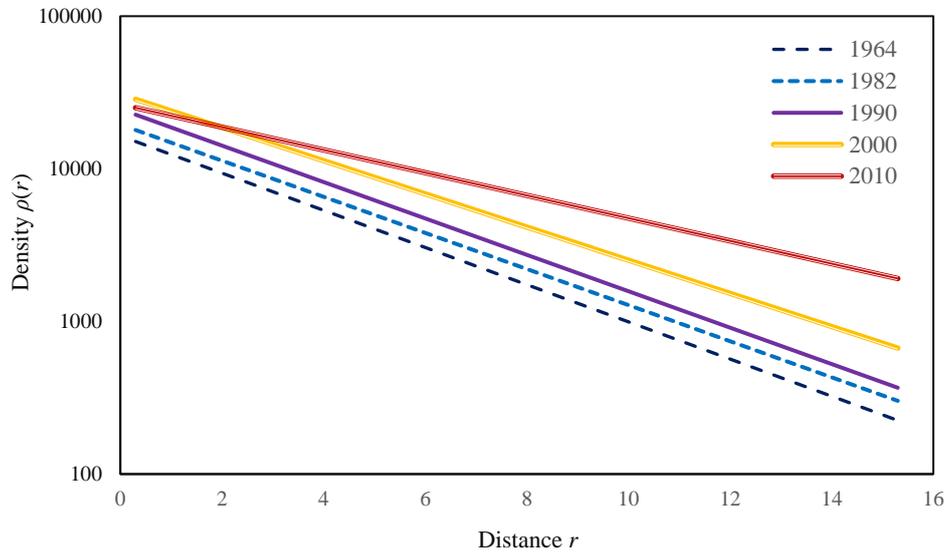

(b)

**Figure 2**. Comparing the gradients of population density decay in Hangzhou in different years (1964-2000). **(a)** Plot of urban density decay; **(b)** Semi-log plot for urban density decay.

## 3.3. The spatial restructuring of population in Hangzhou

Now, it is necessary to discuss the spatial restructuring of population in Hangzhou city by analyzing zonal datasets. The data analysis based on the zonal system consisting sub-districts with irregular shape differs from that based on the concentric ring system with regular shape. Data analysis based on concentric circles can reflect the global statistical law and core-periphery relationship of urban growth, while the data analysis based on sub-districts can mirror the local variation feature of spatial distribution. Comparing the figures of the spatial distribution of population density by sub-districts in Hangzhou in 1964, 1982, 1990, 2000, and 2010, we can find that the tendency of concentration of population is weakened (Figure 3). On one hand, the number of the largest population density of sub-district decreased from 62,596 p/km$^2$ in 1964, to 57,977 p/km$^2$ in 1982, to 47,369 p/km$^2$ in 1990, to 33,707 p/km$^2$ in 2000 and then to 46738 p/km$^2$ in 2010; on the other hand, the number of the sub-districts with population density more than 10,000 p/km$^2$ increased, and the extent of such sub-districts was enlarged, for example, in 1964 and 1982 most of them were distributed within 4.8 km from the city center, while in 1990, 2000 and 2010 most of them were distributed within 6.2 km from the city center.

The spatial reconstruction of a city can be mirrored by the spatio-temporal variation of population density. The spatial distribution of the growth rate of population density in 1964-1982, 1982-1990, 1990-2000 and 2000-2010 shows that there are four stages in the process of spatial restructuring of population in Hangzhou in about half a Century since 1964 (Figure 4). In the first stage, the rural population increased, and then the population in the core area grew slowly. In the second stage, the population in the core area began to decline, and the population in the inner suburb began to concentrate and increase clearly. In the third stage, the population of the core area decreased fast, while that of the inner suburban area increased rapidly. In the fourth stage, the population of the core area continued decrease, while that of the inner suburb kept increase. In fact, there were two forces in the spatial restructuring of population in Hangzhou. One is the centrifugal force resulting from the decentralization of population in the core, and the other is the centripetal force resulting from the centralization of population (Feng and Zhou, 2005). The changes of sub-district population density tell us that suburbanization in Hangzhou occurred in 1982-1990 and accelerated in the 1990s and in the 2000s, and that the period of 1964-1982 may be the prelude of decentralization of population.

The features of urban growth can be reflected by two aspects. One is the relationship between the growth rate of population densities and the distance from the city center, and the other is the relationship between the growth rate of population densities and densities of the former year in a certain period. The growth rate of population density was limited in the period 1964-1982, while that in the period 1982-1990, 1990-2000 and 2000-2010 was larger. For example, the largest growth rate of population density in 1964-1982, 1982-1990, 1990-2000 and 2000-2010 was 112.8 percent, 717.7 percent, 449.3 percent and 388.9 percent, respectively. For another example, as far as the proportion of the sub-districts with growth rate over 100 percent to total sub-districts is concerned, it was 6.3 percent, 12.5 percent, 23.5 percent and 22.2 percent in 1964-1982, 1982-1990, 1990-2000 and 2000-2010, respectively. Compared with population density in

1964-1982, more and more sub-districts within the distance of 5 km from the city center began to lessen density in 1982-1990, 1990-2000 and 2000-2010, while sub-districts with the distance of 5 to 12 km from the city center gained more and more density especially in the period 1990-2000 and 2000-2010, and some sub-districts with the distance of 12 to 20 km from the city center gained density clearly in the period 2000-2010 (Figure 4a). On the other hand, it can be found that the sub-districts with larger growth of population density were concentrated in the position of lower population density especially less than 5,000 p/ km$^2$. Compared with that in the period 1964-1982, some of the sub-districts with population density less than 5,000 p/ k m$^2$ gained more density in the period 1982-1990, 1990-2000 and 2000-2010 (Figure 4b). On the countryside, many sub-districts of the ones with population density more than 25,000 p/km$^2$ lost their densities in the period 1982-1990, 1990-2000 and 2000-2010.

As far as the sub-districts in 1964, 1982, 1990, 2000 and 2010 are concerned, no matter they are near or far from the city center, and no matter the area of these sub-districts (i.e. patches on the map) is large or small, the absolute value of population density is reflected in the figures. Basically, it can be seen that the sub-districts with the highest population density are not in the place of the city center, but in the places that are a certain distance from the city center, and not too far away from it (Figures 3(a) and (b)). The change in the percentage of population density in the sub-districts (Figures 4(a) and (b)) reflects the growth of the population living at different distances from the city center, which can reflect the actual development of the region in Hangzhou. For example, the suburbanization of the central area makes the population density of the sub-districts in the central area negative growth, and the real estate development near the inner suburbs makes the population of buying houses in this area increase greatly, which leads to the growth of population density of sub-districts in this area. It is worth pointing out that there are differences in the growth of population density in different periods. For example, during the period 1990-2000, the sub-districts with obvious population growth mainly concentrated in the inner part of the inner suburb, while during the period 2000-2010, this situation changed, and the sub-districts with obvious population growth evolved into the outer part of the inner suburb, which indicates that after 2000, the suburbanization of Hangzhou continued and has gone further. This phenomenon indicates the spatial re-organizing process of decentralization and concentration of urban population in different regions. As we know, population density is inversely related to the distance from the city center. For example, sub-districts in the core with shorter distance from the city center reduced their densities in the 1980s, the 1990s and the 2000s, indicating they played important roles in population decentralization. For another example, some sub-districts in the outer part of the inner suburb with longer distance from the city center, increase more densities in the 1990s and the 2000s, indicating that such areas played important roles in population concentration.

### 3.4. Suburbanization accounts for the spatial restructuring of population

Spatial restructuring process of urban population is a spatial dynamics of population migration. Suburbanization is closely related to the changes of urban internal spatial structure in Hangzhou (Feng and Zhou, 2005). The modeling results show that the gradient in the negative exponential function of Hangzhou tended to be flatter especially since the 1990s. The actual values of population density of the city center declined. The above facts should be attributed to the development of suburbanization. Suburbanization is associated with the dynamics of increasing of information entropy of population distribution in Hangzhou. It is suburbanization and decentralization that weakened the difference of population distribution between the core and its periphery. It should be pointed out that suburbanization in Hangzhou was still at an incipient stage before 2000. As a result, the theoretical value of population density of the city center in the Clark's model still keeps increasing. However, it decreased in the period 2000-2010, indicating that suburbanization in Hangzhou in the new century has entered a new stage of development, which is like what happens in the Western cities.

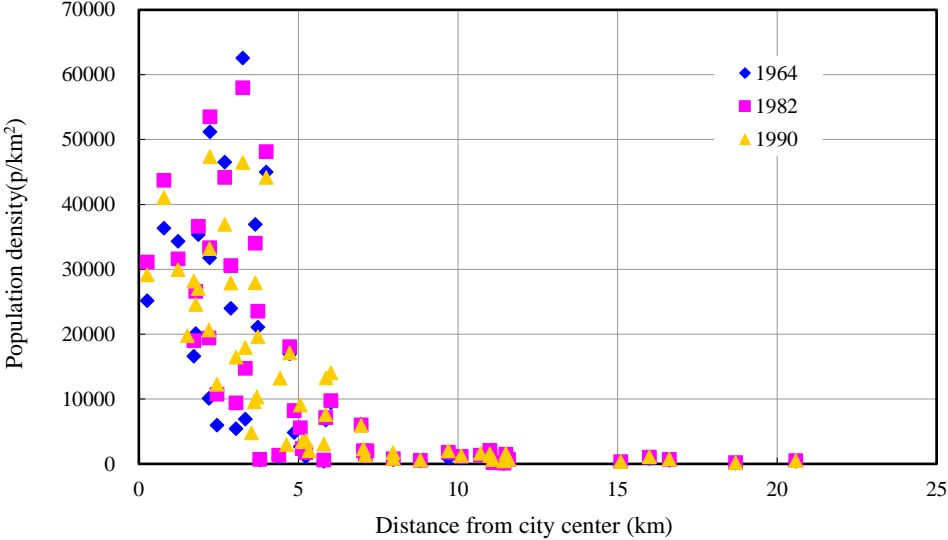

(a)

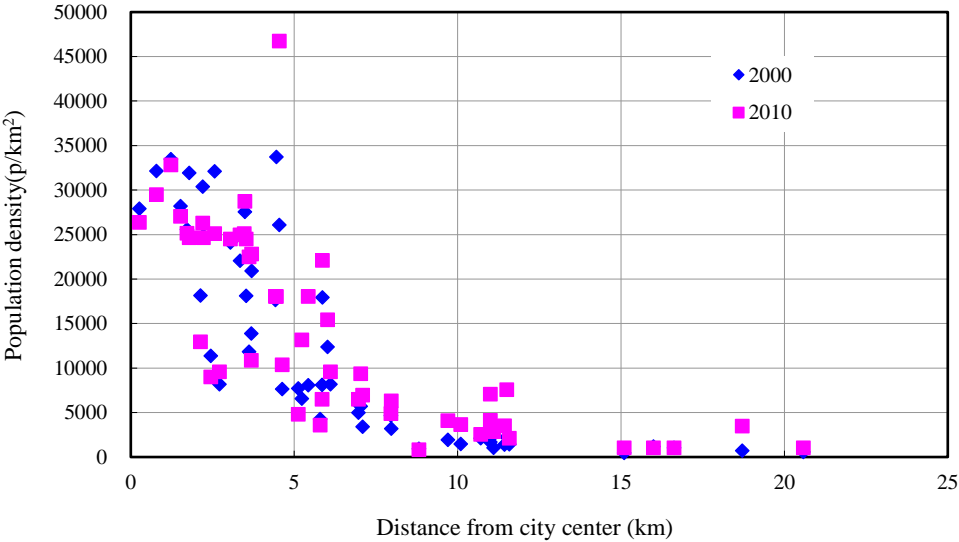

(b)

**Figure 3**. Population density by sub-district vs distance from the city center in Hangzhou (1964, 1982, 1990, 2000 and 2010). **(a)** Population density vs distance from the city center in 1964, 1982, and 1990; **(b)** Population density vs distance from the city center in 2000 and 2010. **Source:** Data from Hangzhou's Census

Office. The data materials are as follows: (1) Manually Tabulated Data of the Second Population Census of Hangzhou, 1964; (2) Manually Tabulated Data of the Second Population Census of Hangzhou, 1982; (3) Manually Tabulated Data of the Second Population Census of Hangzhou, 1990; (4) Manually Tabulated Data of the Second Population Census of Hangzhou, 2000; (5) Manually Tabulated Data of the Second Population Census of Hangzhou, 2010.

The causes of Hangzhou's suburbanization are summarized as follows:

(1) The development of transportation facilities and fast growing of automobiles. Since the 1980s, the performance of the reformation and opening-up policy changed the way of investment and construction of Hangzhou's traffic from state monopoly to the multi investment system. The former was based on traditional planned economy, and the latter was based on joint venture and commodity economy. The construction of urban traffic improves the relationship between the urban core and the suburbs. For example, the first subway in Hangzhou was put into use in 2012, and so far five metro lines have been constructed. Most of these subway lines play the role of effectively connecting the central area and the suburbs, facilitating the long-distance commuting of citizens, and effectively promoting the suburbanization of residence. The fast growing of automobiles also propelled the development of suburbanization in Hangzhou since the 1990s. For example, the number of automobiles increased 908 thousand from 170 thousand in 1996 to 1,078 thousand in 2005, with the growth rate of 532.9 percent and the annual growth rate of 22.8 percent (Hangzhou's Statistics Bureau, 1997; Hangzhou's Statistics Bureau, 2006). The private automobile came into urban families in Hangzhou so that some of the driving forces of suburbanization are similar to the Western cities.

(2) The renovation of the core and the construction of new residential quarters in suburbs. The large scale renovation of the old city began in Hangzhou in the early 1980s. In 1986, the government of Hangzhou put forward that urban development should combine the renovation of the old city with the construction of new quarters in suburbs. In the period from 1986 to the first half year of 1999, as the result of rehabilitation, houses totaling 8.75 million sq m living space were demolished, and 110, 000 households and work units (*danwei*) had to move away from the core. In the course of renovation of the core, most of the household whose homes were demolished were relocated in the suburbs. At the same time, about 130 new residential quarters, which are attractive to urban residents because of their attractive environments, facilities, and cheap prices, were constructed in the inner suburb (Feng and Zhou, 2005). Accordingly, residential suburbanization was promoted.

(3) The reform of urban land use system and the spatial pattern of land and housing price. In the planned economy, urban land values in China were not evaluated and land was charged with a small fixed rate (Wang and Zhou, 1999; Wang and Meng, 1999). Since the year of 1992, with the establishment of the system of paid urban land use, the market of urban and real estate have been developing in Hangzhou. This change of urban land use system brought the shift of land use in the core from industrial to commercial and other tertiary uses (Zhou and Ma, 2000). In order to reduce the cost and obtain more space, more and more factories and their employees moved out of the core to the suburbs. Development of land market acted on the spatial patterns of the prices of land and housing in Hangzhou, which decay from the core to the peripheral area. This pattern of housing prices guides the development of suburbanization and decentralization of population. With the fast growing of housing price, more and more housing purchasers choose housing in suburbs in order to save money or to buy more space with the same money.

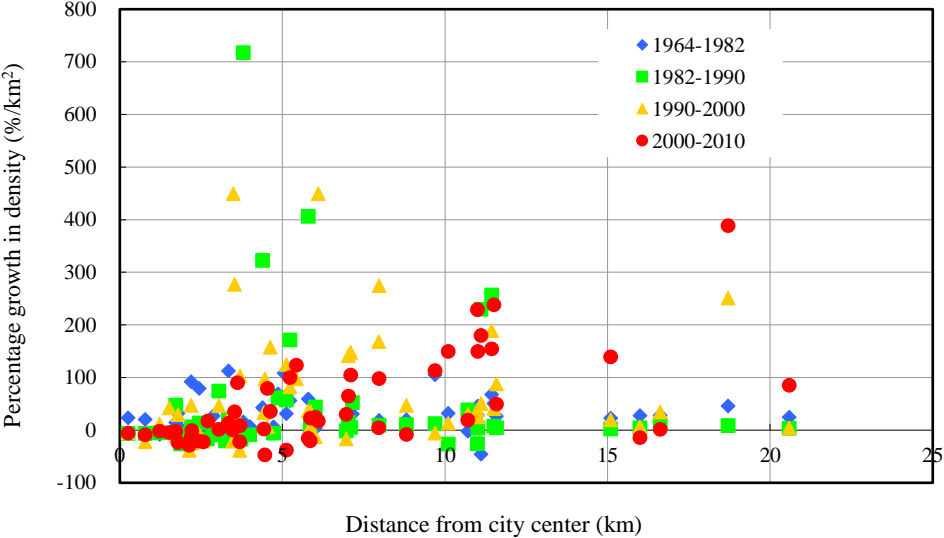

(a)

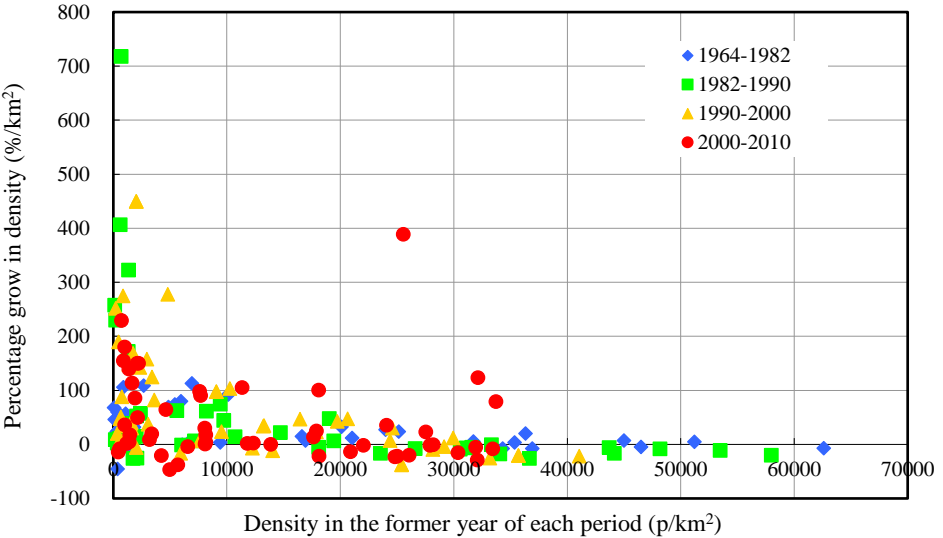

(b)

**Figure 4.** Restructuring of population density in Hangzhou (1964-1982, 1982-1990, 1990-2000 and 2000-2010). **(a)** The percentage growth in population density vs the distance from the city center; **(b)** The percentage growth in population density vs density in the former year of each period. **Source**: Data from Hangzhou's Census Office (1964, 1982, 1990, 2000 and 2010).

### 3.5. The changes of economic system and urban growth

The information entropy of Hangzhou's population distribution in 1964-2010 kept increasing but fluctuated. This indicates that the urban geographic system of Hangzhou is becoming more complex and orderly than ever, albeit the information entropy decreased in the period 1964-1982. Since the 1980s, the policy of reformation and opening-up and the establishment of market economy exerted tremendous influences on urban growth and the spatio-temporal structure of Hangzhou's population density. The policy of reformation and opening-up made the city of Hangzhou become an opening system, while the establishment of market economy strengthened the ability of Hangzhou's urban system as a self-organization. Although parts of a self-organization are not in order, the whole of it can evolve into highly ordered structure. It is no doubt that the performance of the reformation and opening-up policy and the transformation from a planned economy to a market one all increased the ability of Hangzhou city as self-organizing system. Order emerges from instability and from random growth spurts (Benguigui *et al*, 2000; Benguigui *et al*, 2001). In a word, self-organizing evolution strengthened the spatial complexity of Hangzhou city, and raised the utilities of Hangzhou's function units. In this way, the total tendency of the changes of information entropy or the parameter σ of the power-exponential model in Hangzhou since the 1980s can be explained.

Economic system may also have something to do with the goodness of fit of the Clark's model to urban density data. As we know, the negative exponential function was demonstrated to describe population densities in the Western cities successfully. It is also applicable to Chinese cities with the market economy or socialist market economy (Wang and Zhou, 1999; Wang and Meng, 1999; Feng *et al*, 2009). However, till now it has not been proved whether the Clark's model is applicable to Chinese cities with the early planned economy especially before the 1980s. In Hangzhou, the fact that the goodness of fit of the negative exponential function was worse than that of the logarithmic function in 1964 and 1982 can be attributed to the early planned economy.

Both the value of information entropy and the parameter σ of the power-exponential model declined during the period from 1964 to 1982. This may be associated with the disorder of urban construction and development in the period of the Cultural Revolution (1966 to 1976). Urbanization during the Culture Revolution was marked by negative sentiments of anti-urbanism and the attendant program of rusticating urban youths, which was a unique episode in China's urban development (Zhou and Ma, 2000). In the period, the Three-Lines (*Sanxian*) Construction entered its peak, with factories transferred from cities to the mountainous areas and distributed dispersedly (Zhou, 1995). Totally speaking, the Culture Revolution made urban China to be in a stage out of order. The properties include anarchism prevailing, investment of urban construction declining, urban planning being abandoned, and lack of urban infrastructure and slower development or degradation of urban economy. In Hangzhou, the fact that information entropy of population distribution declined in the period 1964-1982 means the chaos of urban development in the period of Cultural Revolution.

## 4. Discussion

The empirical analytical results show that, based on the statistical average processing of spatial data, Hangzhou's population density can be modeled by two functions. One is negative exponential function, and the other is power-exponential function. The former is termed Clark's model in literature (Batty and Longley, 1994; Clark, 1951), and the latter is the generalized results from Clark's model and Sharratt's model (Chen, 2010). Model is different from truth. The result of mathematical modeling of a complex system is not unique. Different types of models have different uses in scientific research. As indicated above, geography has two worlds: real world and ideal world (Tang, 2009). On the other hand, system research falls into three categories: behavioral research, values research, and normative research (Krone, 1980). Clark's model is derivable from entropy maximization principle, indicating system optimization process (Batty and Longley, 1994; Chen, 2008). Therefore, Clark's model can be used to model the ideal state of Hangzhou's population distribution for normative research. In contrast, the power-exponential model can be used to model the real state of Hangzhou's population distribution for behavioral research, that is, empirical studies (Table 5). The latent scaling exponent values reflect the extent of deviation of real state to optimized state of urban population distribution. The gap between the real state and ideal state is just the problem to be solved by city planning and design. For system analysis, the gap between the goal and the status quo is the problem to be solved (Forester, 1969; Su, 1988). By comparing the differences between the characteristic radius based on Clark's model and that based on the power-exponential model, we can judge the distance between the real state and expected state of urban population distribution. The above analysis shows that, based on good social and economic environment, a city will automatically tend to the optimal state through the self-organization process. It lends further support to the viewpoint that geographical laws of urban systems are the law of evolution rather than the law of existence (Chen, 2016).

Table 5. Comparison for two models of Hangzhou's urban population density distribution

| Geographical world | Research type | Model | Model type | Modeling method | Parameter |
|---|---|---|---|---|---|
| **Real world** | Behavioral research | Power-exponential model | Parameter model | Analytical and experimental method | Real characteristic radius $r_0$ and latent scaling exponent $\sigma$ |
| **Ideal world** | Normative research | Negative model (Clark's model) | Mechanism model | Experimental model | Ideal characteristic radius $r_{0*}$ |

**Note**: Clark's model is originally a parameter model based on empirical analyses. After it was derived from the principle of entropy maximization, the model had become a mechanism model with clear theoretical meaning.

By mathematical modeling and empirical analysis, we obtained new insight into Hangzhou's development from 1964 to 2010. The spatio-temporal evolution of Hangzhou has different influencing factors and realistic characteristics at different stages (Table 6). Before 1982, the driving force of urban development was the simple top-down force based on command economics, while after 1982, the dual force based on both the top-down government management and bottom-up market economics influence urban growth. New changes of both external environment and internal factor in Hangzhou after 1982 include the development of transportation facilities and fast growing of automobiles, the renovation of the core and the construction of new residential quarters in suburbs, and the reform of urban land use system and the spatial pattern of land and housing price. The resultant of forces of these changes derived suburbanization and decentralization of population in the city. From 2000 to 2010, real estate and city building movement began to affect urban development. In particular, real estate has left a significant mark on the urban population distribution of Hangzhou.

Table 6. Internal and external factors affecting the evolution of urban population distribution in Hangzhou in different stages

| Stage | External environment | Internal factor |
|---|---|---|
| 1964-1982 | Political movement (Cultural Revolution); planned economy; Lack of urban planning; The background of unit society (Danwei) | Large scale construction of unit compound in unit system society; The residential area of the unit built in a centralized way; Decentralized distribution of urban industry |
| 1982-1990 | Reform and opening-up policy; After the reform and opening up, begin to work out the urban master plan for the first time and play a guiding role in urban construction and development; Urban development following the idea of centripetal agglomeration | The relocation of residents caused by the transformation of urban center and old city under the guidance of the government; Residents forced to move due to the construction of transportation facilities; Under the effect of the land price difference between urban and rural areas, some industries began to move out |
| 1990-2000 | Reform and opening-up policy, socialist market economy; Establishment of the system of paid use of land; Reform of housing system; The idea of urban development is gradually changing from the idea of agglomeration to that of centrifugal diffusion | The construction of a large number of bus lines has improved the connection between urban and rural areas; A large number of high-speed highway construction; A large number of industrial enterprises move out; The development of private cars; The construction of suburban affordable housing community; Construction of suburban residential area; Construction of villas in suburbs; The rise of suburban shopping centers |
| 2000-2010 | Market economy, real estate development; Implementation of the policy of invigorating the city through industry; The development concept from the West Lake era to the Qiantang River Era; Implementation of new town construction policy; Implementation of Development Zone Construction Policy | The rapid development of private cars; Large scale construction of residential areas in suburbs; Suburban shopping centers at all levels were popularized; In 2007, the subway began planning and construction and put into use five years later; Active suburbanization to improve living area and living environment; he rise and development of suburban development zone makes the employment suburbanization |

This research makes contributions to the literature in the following aspects. Firstly, it provides an empirical research on population densities in another important Chinese city, Hangzhou, an ancient capital of China. As the above-mentioned, less research on population density is undertaken in China. As for the reasons, one is that data are less plentiful and less reliable in developing countries, and the other is that the scarcity of public accessible research data on China is evident because of the country's longtime concern of national security and reluctance of releasing data to the public (Mills and Tan, 1980; Wang and Meng, 1999). Moreover, an important reason comes from the frequent adjustment of administrative divisions by Chinese local government, who often ignore providing convenience for comparative research of one district or sub-district in different year. For example, the geographical database of China's census, called sub-district, often changes over time. As a result, clarifying the evolution of the boundaries of sub-district and deal with sub-district data should be important tasks in studying urban growth and population densities in China. Fieldwork was carried out in Hangzhou to obtain unpublished data of population of sub-districts and to ascertain the spatial boundaries of the sub-districts in Hangzhou from 1964 to 2010. Secondly, it tries to indicate the development characters of urban population densities in China in the late stage of reform and opening up (1990-2000) and the ten years in the new century. It is important to note that the previous literatures on population densities in China used data from the third (in 1982) and the fourth (in 1990) censuses. Therefore, we do not know much about that in the 1990s and the 2000s. Only because data from the fifth census (in 2000) and the sixth census (in 2010) became available has such a study been possible. Thirdly and most important of all, it tries to prescribe the feature of urban growth and the spatial restructuring of population in Hangzhou in a long period from 1964 to 2010. In the past half a century before 2010, China's urbanization experienced complex changes, caused by multiplex driving forces. The driving forces include socialist planned economy in the period 1960-1965 and the period of the 1980s, the Cultural Revolution (1966-1976) marked by unprecedented political movement and an retrogression of economy in the nation's history (Zhou and Ma, 2000), and socialist market economy or transitional market economy in the 1990s and 2000s.

Unlike the West where urban population density has attracted a great deal of geographers' attention, research on population density distribution in urban China is still scarce although since the end of the 1990s some research progress has been made. This research provides the literature with an interesting case on population densities in another important socialist city, Hangzhou, an ancient capital of China. Moreover and most important of all, modeling urban population densities also provides us an effective tool to indicate the features of urban growth and spatial restructuring of population. The chief shortcomings of this study are as follows. Firstly, due to the lack of necessary spatial data, the distribution and evolution of urban population density after 2010 are not analyzed. As mentioned above, the pattern of administrative divisions in China has been constantly adjusted. The zonal system of sub-districts in Hangzhou city change continuously. Only by obtaining both the change map of Hangzhou Street (sub-districts) boundary and the data of the seventh census (2020), can we transform it into the data of urban population density distribution and make analyses for recent years' change. Secondly, limited to the length of the paper, there is no computer simulation of urban evolution mechanism. Computer simulation technology can make up for the lack of experiment in urban research, and help us to deeply analyze the mechanism of urban development (Batty *et al*, 1989; Fotheringham *et al*, 1989; Tobler, 1970; White and Engelen, 1993). The work will be carried out in the future.

# 5. Conclusions

Urban study should proceed first by describing how an urban phenomenon and its parts work and then understanding why. Description depends on mathematics and measurement, and understanding relies heavily on observation, experience, artificially constructed experiments, and computer simulation. The premise of understanding is effective description, while performance is convincing explanation. The main work of this paper lies in two aspects: describing population density distribution of Hangzhou city and explain its antecedents and consequences.

One is the description of urban growth by modeling population density distribution. Seven functions are tested with the data of Hangzhou in 1964, 1982, 1990, 2000, and 2010, respectively. The result shows that the negative exponential function and the power-exponential one occupy significant positions in modeling population densities of Hangzhou. In fact, these two functions have close relationship with each other, for the former is the ideal form of the latter. With the aid of the changes of some parameters in the regressive models, we try to analyze the features of urban growth in Hangzhou in the last forty years of the 20th Century. The change of the parameter σ in the power-exponential function reflects the tendency of change of information entropy of urban geographic system in Hangzhou. The parameter values kept increasing but fluctuated, indicating that the features of urban growth characterized by information entropy-maximizing and self-organizing. In the period 1964-1982, the information entropy decreased showing a degradation of urban economy and construction in Hangzhou, which should be attributed to the Culture Revolution (1966-1976). Since the 1980s, the policy of reformation and opening-up and the establishment of a market economy exerted tremendous influences on urban growth. Reformation and opening-up made Hangzhou become an opening system, and market economy strengthened the ability of Hangzhou's urban system as a self-organization, which is a behavior from below to top.

The other is the explanation of urban evolution in Hangzhou by analyzing growing factors. The changes of the gradient in the negative exponential function of Hangzhou have something to do with the concentration and decentralization of population. In the past half a century before 2010, the process of spatial restructuring of population in Hangzhou can be divided into three stages. The features of the three stages are as follows: (1) population development in rural area and progressively slower increase of population in the core in the first stage; (2) population decline in the core and population concentration in part of the inner suburb in the second stage; and (3) fast population decline in the core and fast population concentration in the inner suburb in the third stage. The gradient becoming flatter in 2000 and 2010 is closely related to the development of suburbanization and decentralization of population in Hangzhou. It is also noticed that Hangzhou's suburbanization was still at an incipient stage before 2000, for the theoretical value of population density of the city center in the Clark's model still keeps increasing although its actual value declined in the 1990s. However, it decreased in the period 2000-2010, indicating that suburbanization in Hangzhou in the new century has entered a new stage of development, which is like what happens in the Western cities. Hangzhou's suburbanization and decentralization of population have their causes, including the development of transportation facilities and fast growing of automobiles, the renovation of the core and the construction of new residential quarters in suburbs, and the reform of urban land use system and the spatial pattern of land and housing price. In a word, both internal and external factors affect the evolution of urban population distribution in Hangzhou in different stages.

**Acknowledgments:** This research was funded by the National Natural Science Foundations of China, grant numbers 41671157 & 41671167.